\documentclass[prd, aps, superscriptaddress, preprintnumbers, twocolumn, floatfix, nofootinbib]{revtex4}
\pdfoutput=1

\usepackage{amsfonts}
\usepackage{amsmath}
\usepackage{amssymb}
\usepackage{bm}
\usepackage{dcolumn}
\usepackage{graphicx}   
\usepackage[latin1]{inputenc}
\usepackage{latexsym}
\usepackage{rotating}
\usepackage{hyperref}
\usepackage{graphicx}
\usepackage{color}

\newcommand\be{\begin{equation}}
\newcommand\ba{\begin{eqnarray}}
\newcommand\ee{\end{equation}}
\newcommand\ea{\end{eqnarray}}

\begin{document}

\title{Global 21cm Absorption Signal from Superconducting Cosmic Strings}

\author{Roxane Th\'{e}riault}
\email{roxane.theriault2@mail.mcgill.ca}
\affiliation{Department of Physics, McGill University, Montr\'{e}al. QC, H3A 2T8, Canada}

\author{Jordan T. Mirocha}
\email{jordan.mirocha@mcgill.ca}
\affiliation{McGill Space Institute and McGill Physics Department, Montr\'{e}al, QC, H3A 2T8, Canada}

\author{Robert Brandenberger}
\email{rhb@physics.mcgill.ca}
\affiliation{Department of Physics, McGill University, Montr\'{e}al, QC, H3A 2T8, Canada}

\date{\today}

%%%%%%%%%%%%%%%%%%%%%%%%%%%%%%%%%%%%%%%%%%%%%%%%%%%%%%%%%%%%%%%%%%%%%%%%%%%%%%%%%%%%%%%%%%%%%%
\begin{abstract}

Superconducting cosmic strings emit electromagnetic waves between the times of recombination and reionization. Hence, they have an effect on the global 21cm signal. We compute the resulting absorption features, focusing on strings with critical current, study their dependence on the string tension $\mu$, and compare with observational results. For string tensions in the range of $G \mu = 10^{-10}$, where $G$ is Newton's gravitational constant, there is an interesting amplification of the two characteristic absorption features, one during the cosmic dawn, $z \lesssim 30$, and the other during the cosmic dark age, $z \sim 80$, the former being comparable in amplitude to what was observed by the EDGES experiment.

\end{abstract}
%%%%%%%%%%%%%%%%%%%%%%%%%%%%%%%%%%%%%%%%%%%%%%%%%%%%%%%%%%%%%%%%%%%%%%%%%%%%%%%%%%%%%%%%%%%%%%

\maketitle

%%%%%%%%%%%%%%%%%%%%%%%%%%%%%%%%%%%%%%%%%%%%%%%%%%%%%%%%%%%%%%%%%%%%%%%%%%%%%%%%%%%%%%%%%%%%%%
\section{Introduction} 
\label{sec:intro}

21cm cosmology is emerging as a promising window to probe the distribution of matter in the early universe, in particular in the ``dark ages'' before the onset of reionization related to the formation of the first stars (see e.g. \cite{Furlanetto} for a detailed review). This window is particularly promising to search for signatures of early universe scenarios in which the distribution of matter at early times contains specific nonlinear features. The reason is that, according to the $\Lambda$CDM model, the standard paradigm of early universe cosmology, the distribution of matter at times before reionization is given by a roughly scale-invariant spectrum of linear perturbations, and nonlinear features stick out very clearly, while once the underlying matter distribution becomes nonlinear (which happens around the time of reionization) the primordial nonlinearities become obscured. 

Specifically, early universe models which contain topological defects such as cosmic strings generate characteristic nonlinear structures in the distribution of matter, and their signals for three dimensional 21cm redshift maps were worked out in \cite{Holder1} (see also \cite{extensions}). Early universe cosmology can also be probed using the {\it global 21cm signal}, the 21cm intensity averaged over the sky as a function of redshift (see e.g. \cite{Furl}). Primordial nonlinear structures such as cosmic strings also lead to signatures in the global 21cm signal. As stressed in \cite{Oscar}, global 21cm observations yield a promising window to probe for the possible existence of cosmic strings.
   
Cosmic strings are topological defects analogous to vortex lines in superconductors and superfluids which are predicted to arise in a large class of particle physics models beyond the Standard Model (see \cite{CSrevs} for reviews on the cosmology of cosmic strings). The key point is that in models which have cosmic strings solutions, a network of strings inevitably will form in the early universe and persist to the present time \cite{Kibble}. The distribution of strings rapidly approaches a {\it scaling} solution in which the statistical properties of the string network is independent of time if all lengths are scaled to the horizon. The network consists of ``long'' strings (strings with curvature radius comparable or larger than the horizon, and a distribution of string loops which have formed from the inter-commutation of the long strings. 

In a large set of models, the cosmic strings are superconducting \cite{Witten}, carrying either bosonic or fermionic supercurrents. Non-superconducting string loops oscillate and decay by emitting gravitational radiation while for superconducting strings electromagnetic radiation is important.The electromagnetic radiation from superconducting cosmic strings leads to an enhanced radio photon background at all times between recombination and the present time, but in particular in the ``dark ages'' before the time of reionization. This radio background, in turn, will lead to an enhanced absorption feature in the global 21cm signal \cite{Gil}, and it is interesting to ask how the resulting signal compares with the recent EDGES results \cite{EDGES}, and in particular whether it can reproduce the sharp absorption trough which the observations indicate \footnote{Note that concerns have been raised related to possible foreground contamination of the EDGES results \cite{concern}. We will return to this issue in the concluding section.}.  

For non-superconducting strings, the amplitude of the induced 21cm absorption signal is determined by the single free parameter of the string network, namely the string tension $\mu$, which is usually expressed in terms of the dimensionless quantity $G \mu$, where $G$ is Newton's gravitational constant. Even for such strings, there is a mechanism which produces a flux of photons, namely {\it string cusp annihilation} \cite{RHBcusp}. However, it was found \cite{us1} that the amplitude of the effect is too small to reproduce the amplitude reported in \cite{EDGES}, independent of the value of the string tension. 

 For superconducting strings the radio photon flux depends both on $G \mu$ and on the current. The overall amplitude of the  induced 21cm absorption feature and its dependence on the two-dimensional parameter space of string current and string tension was studied in \cite{us2}. It was found that the effect can be important, and that a region in the two-dimensional tension-current parameter space can be ruled out because the absorption signal would be larger than what \cite{EDGES} has seen. 
 
In this paper we take a closer look at the predicted redshift-dependence of the predicted absorption signal, considering various models for the X-ray heating at lower redshifts. We find a sharp increase in the magnitude of the absorption signal coming from the high redshift side, a feature similar to what is seen in \cite{EDGES}, but the decrease of the signal towards lower redshifts is more gradual than what was reported in \cite{EDGES}. We will here focus on strings with critical current.
 
In this paper we use natural units in which the speed of light, Planck's constant and Boltzmann's constant are set to $1$. The physical time is denoted by $t$, with $t_0$ being the present time and $t_{eq}$ being the time of equal matter and radiation. The corresponding radiation temperatures are denoted by $T, T_0$ and $T_{eq}$, respectively. Instead of time we will often use cosmological redshift $z(t)$. Newton's gravitational constant is denoted by $G$, and $m_{pl}$ denotes the Planck mass.

\section{Brief Review of Superconducting Cosmic Strings}

Cosmic strings are linear topological defects which arise in gauge field theories in which the manifold of ground states has non-vanishing first homotopy group, i.e. a vacuum manifold which ``looks like'' a circle. After a phase transition in the early universe, a network of strings forms. The network rapidly approaches a ``scaling solution'' characterized by a random walk-like set of infinite strings with curvature radius comparable to the horizon, and a distribution of string loops which result from the intersection of the long strings. At times later than the time $t_{eq}$ of equal matter and radiation, the number density of loops per unit radius $R$ is given by \cite{onescale}
\be \label{looplow}
n(R, t) \, = \nu R^{-5/2} t_{eq}^{1/2} t^{-2} 
\ee
for loops formed before $t_{eq}$ (i.e. $R < \alpha t_{eq}$, where $\alpha$ is a constant which characterizes the radius at the time of loop formation), and
\be \label{loophigh}
n(R, t) \, \sim \, R^{-2} t_{-2} 
\ee
for loops formed later. The constant $\nu$ is proportional to the mean number of long strings per horizon volume, and the choice $\nu = 10$ which we use below is motivated by the results of numerical cosmic string evolution simulations \cite{CSsimuls} \footnote{Note that some field theory simulations of cosmic string evolution \cite{Hind}, while agreeing with the scaling distribution of long strings, do not yield such a scaling distribution of string loops because the loops in these simulations mainly decay into particles.}. Note that the probability that a loop is exactly circular is zero. The mean length of a loop with mean radius $R$ is parametrized as $l = \beta R$, and we choose the value $\beta = 10$ in our study. Note that there is a lower cutoff to the loop distribution: loops with radius smaller than some critical value $R_c$ live for less than one Hubble expansion time, and their effects are negligible.

Non-superconducting strings decay primarily by emitting gravitational radiation. The power of this radiation from a string loop of radius $R$ is
\be
P_g \, = \, \gamma G \mu^2 \, ,
\ee
where $\gamma$ is a constant which is again determined by simulations of gravitational radiation from an oscillating loop, and the value $\gamma = 10$ is chosen below based on the work of \cite{GW}. If gravitational radiation is the dominant loop decay mechanism, then the loop cutoff radius at time $t$ is
\be \label{cutoff}
R_c \, = \, \gamma \beta^{-1} G \mu t \, .
\ee

Cosmic strings which arise in many theories beyond the Standard Model of particle physics are superconducting. They either admit trapped bosonic or fermionic zero modes \cite{Witten} which leads to superconductivity. Currents are set up on these strings from the initial fluctuations of the fields in the early universe. For superconducting strings there is a second decay mechanism for string loops - electromagnetic radiation. The power in this radiation averaged over a loop oscillation period is \cite{Witten2}
\be
P_{em} \, = \, \kappa I \sqrt{\mu} \, ,
\ee
where $I$ is the current on the string and $\kappa$ is a numerical constant of the order one.

There is a critical current $I_c$ above which electromagnetic radiation dominates over graviton emission. This current is given by
\be
I_c \, = \, \kappa^{-1} \gamma (G \mu)^{3/2} m_{pl} \, .
\ee
Note that the energy loss due to cusp annihilation \cite{RHBcusp} is negligible for the large tensions and currents which we consider. If $I > I_c$, then the effective cutoff radius $R_c$ in the loop distribution is determined by the current $I$, and it is no longer given by (\ref{cutoff}).

The radio photon energy density at time $t$ due to the emission from superconducting strings was worked out in detail in \cite{us2}. It is determined by integrating the power of electromagnetic radiation from all string loops present at all times $t^{\prime}$ between the time $t_{rec}$ of recombination and the time $t$, and taking into account the redshifting of number density and frequency of photons between $t^{\prime}$ and $t$. It was found that the emission is dominated by the smallest loops which live for more than one Hubble time, i.e. for loops with $R \sim R_c$. For currents $I < I_c$ the resulting energy density scales as $I^2$, while for $I > I_c$ it scales as $I^{5/6}$.  In this paper we are interested both in the shape of the absorption feature and in the overall amplitude. We will consider strings with critical current, and comment at the end on the changes if the current is different from the critical one.

As derived in \cite{us2}, the contribution of superconducting cosmic strings with critical current $I_c$ to the energy density in radio photons in the frequency interval $[0, \omega_{21}]$ at time $t$ is
\be \label{CSenergy}
\rho_{21}(t) \, = \, 18 \kappa^2 {\tilde{\kappa}} \nu G^{-1} \omega_{21}^{1/3} t_{eq}^{1/2} t^{-2} \gamma^{5/6} \beta^{7/6} (G \mu)^{11/6} t^{-1/6} \, ,
\ee
where ${\tilde{\kappa}}$ is $\kappa$ multiplied by a numerical factor of order one. This is the formula which we will use in the following section to work out the global 21cm absorption signal.

\begin{figure}[htbp]
\centering
\includegraphics[scale=0.5]{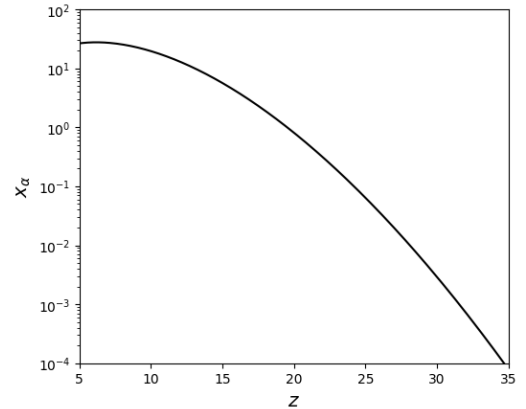}
\caption{Lyman-alpha coupling coefficient as a function of redshift used in this work. This model assumes a 2\% star formation efficiency in atomic cooling halos and adopts a stellar spectrum representative of Population II stars. The coupling coefficient $x_{\alpha}$ (vertical axis) is plotted as a function of redshift (horizontal axis). This is the function used to generate both Figs. 2 and 3.}
\end{figure}

\section{Global 21cm Signal from Superconducting Strings}
 
Let us recall \cite{Furlanetto} that the 21cm  brightness temperature at an observed frequency $\nu$ is given by
\be
T_b(\nu) \, = \, \frac{T_b^{\prime} (\nu_0)}{1 + z} \, ,
\ee
where $\nu_0$ is the frequency of the 21cm line,
\be \label{redshift}
\nu \, \equiv \, \nu(z) \, = \, \frac{\nu_0}{1 + z} \, ,
\ee
and $T_b^{\prime}$ is the brightness temperature at the redshift $z$. The latter is determined by the combination of absorption by the hydrogen gas and emission from the gas, and is given by
\be
T_b^{\prime}(\nu) \, = \, T_s \bigr(1 - e^{- \tau_{\nu}} \bigl) + T_R^{\prime}(\nu) e^{-\tau_{\nu}} \, ,
\ee
where $\tau_{\nu}$ is the optical depth, $T_S$ is the spin temperature of the gas, and $T_R^{\prime}$ is the brightness temperature of the CMB background at the redshift corresponding to $\nu$ (see (\ref{redshift})). Hence
\be
T_b(\nu(z)) \, = \, \frac{1}{1 + z} \Bigr[ T_s \bigr(1 - e^{- \tau_{\nu}} \bigl) + T_R^{\prime}(\nu) e^{-\tau_{\nu}} \Bigl] \, .
\ee
Thus, the relative brightness difference is given by
\ba \label{deltaT1}
\delta T_b(z) \, &=& \, \frac{T_S(z) - T_{\gamma}(z)}{1 + z} \bigr( 1 - e^{-\tau_{\nu_0}} \bigl)
\nonumber \\
&\simeq& \, \frac{T_S(z) - T_{\gamma}(z)}{1 + z} \tau_{\nu_0} \, .
\ea

The optical depth in the early universe is given by \cite{Furlanetto}
\be \label{depth}
\tau_{\nu_0} \, \simeq \, 0.0092 (1 + \delta) (1 + z)^{3/2} \frac{x_{HI}}{T_S} 
\bigr[ \frac{H(z) / (1 + z)}{dv_{\l} / dr_{\l}} \bigl] \ ,
\ee
where $\delta$ is the fractional overdensity of baryons, $x_{HI}$ is the fraction of neutral hydrogen, and $dv_{\l} / dr_{\l}$ is the gradient of the proper velocity along the line of sight.
At the high redshifts we are interested in we have $x_{HI} = 1$. Inserting (\ref{depth}) into (\ref{deltaT1}) then yields
\be \label{deltaT2}
\delta T_b(z) \, \simeq \, [9 {\rm mK}] (1 + z)^{1/2} \bigr(1 - \frac{T_{\gamma(z)}}{T_S} \bigl)
\, ,
\ee
where $T_{\gamma}(z)$ is the equivalent temperature of the photons including both the original CMB photons and those emitted by the superconducting string network.

From \cite{Furlanetto} we recall that the spin temperature can be replaced in terms of the kinetic and color temperaturs $T_K$ and $T_C$ via the relation
\ba
\bigr(1 - \frac{T_{\gamma}}{T_S} \bigl) \, &=& \, 
\frac{x_c}{1 + x_c + x_{\alpha}} \bigr( 1 - \frac{T_{\gamma}}{T_K} \bigl) \, \nonumber \\
&+& \, \frac{x_{\alpha}}{1 + x_c + x_{\alpha}} \bigr( 1 - \frac{T_{\gamma}}{T_C} \bigl) \, ,
\ea
where $x_c$ and $x_{\alpha}$ are the collision and UV scattering coeffiicients, respectively.

To obtain an analytical estimate of the brightness temperatures for redshifts smaller than $z = 25$ we will set $x_c = 0$. If reionization is not yet significant, then $T_C = T_K$ and
\be
T_K \, \simeq \, 0.02K (1 + z)^2 \, .
\ee
In this case, (\ref{deltaT2}) becomes
\be \label{deltaT3}
\delta T_b(z) \, \simeq \, [9{\rm{mK}}] (1 + z)^{1/2} \frac{x_{\alpha}}{1 + x_{\alpha}} 
\bigr(1 - \frac{T_{\gamma}(z)}{0.02K (1 + z)^2} \bigl) \, .
\ee

\begin{widetext}

\begin{figure}[htbp]
\centering
\includegraphics[scale=0.45]{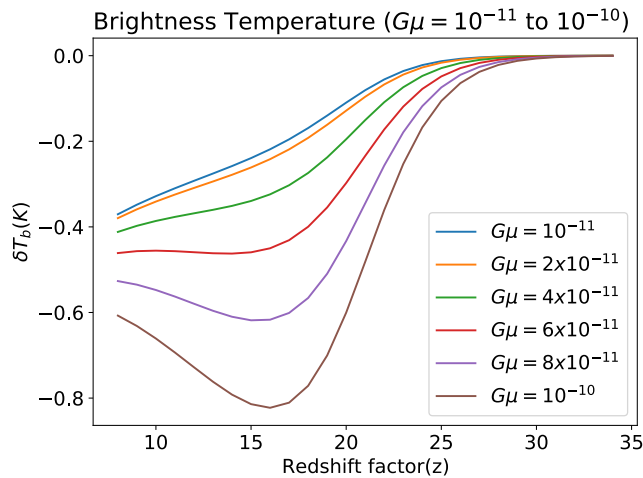}
\caption{The global 21cm absorption signal as a function of redshift for various values of the string tension. The vertical axis gives the brightness temperature signal, the horizontal axis is redshift. To obtain these results, the collision coefficient $x_c$ was set to zero, and the coefficient $x_{\alpha}$ was taken to be the one from the previous figure. The values $\nu = \beta = \gamma = 10$ were used.}
\end{figure}

\end{widetext}
 
The photon temperature at the 21cm wavelength is determined in terms of the photon energy density by \cite{us2}
\be \label{CStemp}
T_{\gamma}(z) \, = \, \frac{3 \pi^2}{\omega_{21}^3} \rho_{21}(z) + T_R(z) \, ,
\ee
where $\rho_{21}(z)$ is the energy density in radio photons in the frequency interval $[0, \omega_{21}]$ due to photons emitted by the superconducting cosmic string network (see below), and $T_R(z)$ is the temperature of the background CMB photons.

The Lyman-$\alpha$ coupling coefficient, $x_{\alpha}$, quantifies the strength of Wouthuysen-Field coupling \cite{Wouthuysen1952,Field1958}, and is directly related to the Lyman-alpha background intensity generated by early galaxies. We consider the model shown in Fig. 1, which is derived from a simple numerical model for galaxy formation that we describe in more detail momentarily. Using this form, we can then evaluate the formula (\ref{deltaT3}) as a function of the string tension. The results are shown in Figure 2 (using the values $\nu = \beta = \gamma = 10$). Note that we have included the contribution of all of the radio photons (both those of the background CMB and those produced by the strings) in determining the equivalent photon temperature (\ref{CStemp}).

We see that for a string tension of $G \mu \sim 10^{-10}$ there is a clear absorption feature in the redshift range $[20 < z < 15]$ which has comparable depth to what is seen in the EDGES results \cite{EDGES}. Larger string tensions give a deeper absorption feature and are hence ruled out for the range of currents considered. Note that the onset of the absorption feature at the high redshift end is quite sharp, but the gradual relaxation of the absorption feature at lower redshifts is less steep than what was reported in \cite{EDGES}.
 
For completeness we now give the expression for the relative brightness temperatue due to the emission from superconducting cosmic strings (neglecting the contribution from the background CMB photons). The energy input from the superconducting cosmic string network leads to a contribution to the effective temperature $T_{\gamma}(z)$ of photons as the 21cm frequency $\omega_{21}$ which in terms of the cosmic string parameters is given by (\ref{CSenergy}). Inserting (\ref{CStemp}) into (\ref{deltaT3}) yields
\be \label{deltaT4}
\delta T_b(z) \, \simeq \, [9 {\rm mK}] (1 + z)^{1/2} \frac{x_{\alpha}}{1 + x_{\alpha}}
\bigr(1 - \frac{A (1 + z)^{5/4}}{0.02K} \bigl) 
\, ,
\ee
with
\be
A \, = \, 54 \pi^2 \nu \gamma^{5/6} \beta^{7/6} (G \mu)^{11/6} T_{eq} 
\bigr( \frac{T_{eq}}{\omega_{21}} \bigl)^{8/3} 
\bigr( \frac{m_{pl}}{T_{eq}} \bigl)^{1/3} (z_{eq} + 1)^{-13/4} \, .
\ee
This shows how the temperature scales as a function of the cosmic string parameters. The amplitude of the brightness temperature trough is an increasing function of the string tension $\mu$. The amplitude also increases if the number of strings per Hubble volume (i.e. $\nu$) grows, and it increases if the string loops become more wiggly (i.e. $\beta$ increases), and as the overall rate of radiation from a given loop (i.e. $\gamma$) grows.

\begin{widetext}

\begin{figure}[htbp]
\centering
\includegraphics[scale=0.6]{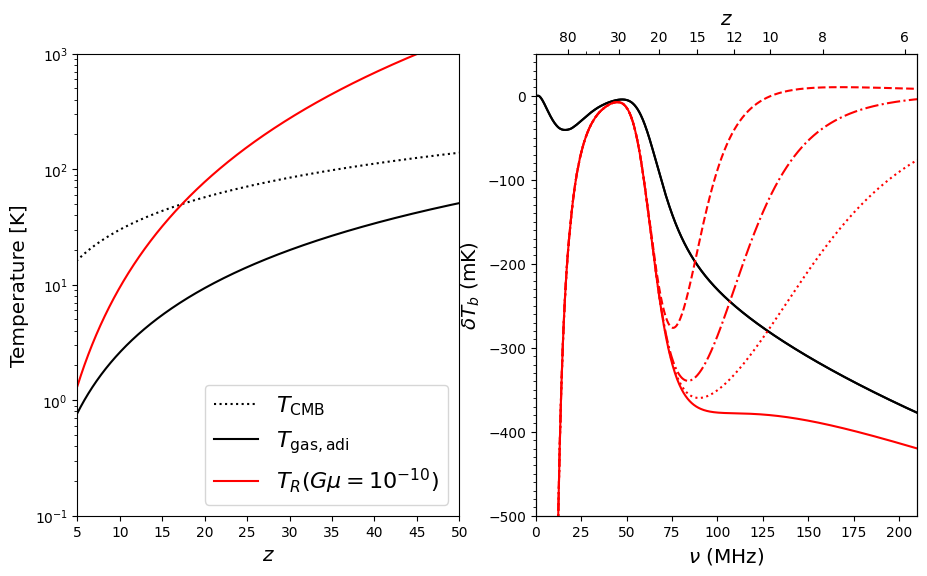}
\caption{The left panel shows the evolution of the CMB temperature $T_{CMB}$, the gas temperature $T_{{\rm gas}}$ assuming adiabatic cooling, and the equivalent 21cm photon temperature $T_R$ which includes the contribution of superconducting cosmic strings with a tension given by $G \mu = 10^{-10}$. The right panel shows the resulting global 21cm signal for this value of the string tension. The solid red curve neglects heating and ionization, the dotted, dash-dotted, and dashed lines steadily increase the efficiency of X-ray heating to $f_X = 0.01, \, 0.1,  \, {\rm and} 1$. The standard $\Lambda$CDM cosmology without strings yields the solid black curve. }
\end{figure}

\end{widetext}

To check these semi-analytical results, we have run several models with the \textsc{ares} code (https://github.com/mirochaj/ares) \cite{code}, which includes the effects of collisional coupling \cite{Zygelman} and computes the radiative coupling using a simple model of galaxy formation. The models presented in this paper were run using the default setup, which relates the cosmic star formation rate density to the rate at which matter collapses into dark matter halos. We assume a star formation efficiency of $f_{\ast}=2$\% in all atomic cooling halos, and assume that 9690 photons are emitted per stellar baryon between Lyman-$\alpha$ and the Lyman limit, as is appropriate for Pop~II stars \cite{Barkana2004}. Note that more recent models allow the star formation efficiency to vary with halo mass \cite{Mirocha2017,Park2019}, and generically predict higher frequency global 21-cm absorption signals. However, for simplicity, here we adopt a more idealized parameterization, and simply tune $f_{\ast}$ to yield a signal that peaks at $\sim 75$ MHz to roughly match EDGES.
 
The results (for a string tension of $G \mu = 10^{-10}$) are shown in the right panel of Figure 3. The red curves are for a model with superconducting cosmic strings, the black curve is the prediction of the standard cosmological model (with $x_{\alpha} = x_c = 0$). The solid red curve is also obtained by setting $x_{\alpha} = x_c = 0$. Comparing the two solid curves shows that the  rather sharp onset of the absorption feature at $z \sim 20$ is a result of the contribution of the strings. The shape of the brightness temperature fluctuation at lower redshifts depends crucially on the model of star formation, i.e. on the redshift dependence of $x_{\alpha}$. The results of three different choices of X-ray heating efficiencies are shown.

\section{Discussion} \label{conclusion}

We have studied the global 21cm signal of superconducting cosmic strings as a function of the string tension. These strings produce a significant flux of radio waves when they decay, and hence lead to a distinctive absorption signal in the 21cm brightness temperature. We found two characteristic features, firstly a sharp absorption trough at a redshift of around $z = 20$, and secondly another absorption trough at higher redshifts. The first absorption trough corresponds to redshifts where the EDGES experiment \cite{EDGES} detected a trough. For a string tension of $G \mu \sim 10^{-10}$, the depth of the trough is similar to what is reported in \cite{EDGES}. The sharp decrease in the relative brightness temperature at the high redshift end of the trough also agrees well with observations, and appears to be a characteristic feature of the cosmic string model. On the other hand, the slope of the brightness curve on the low redshift end of the trough is more gentle than what is reported in \cite{EDGES}, and the slope also depends sensitively on the model of early star formation used.

In our calculations we have assumed that the current $I$ on the superconducting string is the critical current $I_c$. Since $\rho_{21}$ scales as $I^2$ for $I < I_c$, then the absorption signal we predict is constant along a curve of fixed 
\be \label{scaling1}
\bigr( \frac{I}{I_c} \bigl)^{12/11} G \mu \,\,\, I < I_c \, ,
\ee
and since $\rho_{21}$ scales as $I^{5/6}$ for $I > I_c$, the absorption signal is the same along curves of constant
\be \label{scaling2}
\bigr( \frac{I}{I_c} \bigl)^{5/11} G \mu \,\,\, I > I_c \, .
\ee
In evaluating (\ref{scaling1}) and (\ref{scaling2}) it is important to keep in mind that $I_c$ depends on $G \mu$.

It is interesting that the critical value of the string tension for which we obtain an absorption feature of similar depth to what is reported by the EDGES collaboration is of the same order of magnitude as the upper bound on the string tension from pulsar timing array measurements \cite{PTA}. 

As already pointed out in \cite{us2}, models which admit superconducting cosmic strings (with a critical current) with a tension greater than $G \mu = 10^{-10}$ are ruled out since they would generate a global 21cm trough deeper than what is consistent with observations.
In fact, if the EDGES results do not hold up under closer scrutiny, and the absorption trough  
turns out to be less deep than current results indicate, then our analysis would be able to rule out a slightly larger range of values of the string tension for these superconducting cosmic string models. As Figure 2 shows, however, the predicted amplitude of our signal depends strongly on the string tension, and hence we would not expect a significant change in the bound from new data.

\section*{Acknowledgement}

\noindent The research of RB at McGill is supported in part by funds from NSERC and from the Canada Research Chair program. RB is grateful for hospitality of the Institute for Theoretical Physics and the Institute for Particle Physics and Astrophysics of the ETH Zurich. He also wishes to thank Oscar Hernandez for emphasizing the potential of the 21cm global signal to probe for the possible existence of cosmic strings, and Anastasia Fialkov for encouragement to study the redshift dependence of the string-induced absorption feature.

\end{document}